\documentclass[aps,prb,reprint,groupedaddress,showpacs,floatfix]{revtex4-1}

\usepackage{graphicx}
\usepackage{dcolumn}
\usepackage{bm}
\usepackage[mathlines]{lineno}

\begin{document}

\title{\hfill {\small Phys. Rev. B {\bf 86} (2012)}\\
       Computational study of the thermal conductivity in
       defective carbon nanostructures}

\author{Zacharias G. Fthenakis}
\affiliation{Physics and Astronomy Department,
             Michigan State University,
             East Lansing, Michigan 48824, USA}

\author{David Tom\'{a}nek}
\email[E-mail: ]{tomanek@pa.msu.edu}%
\affiliation{Physics and Astronomy Department,
             Michigan State University,
             East Lansing, Michigan 48824, USA}

\date{\today} 

\begin{abstract}
We use non-equilibrium molecular dynamics simulations to study the
adverse role of defects including isotopic impurities on the
thermal conductivity of carbon nanotubes, graphene and graphene
nanoribbons. We find that even in structurally perfect nanotubes
and graphene, isotopic impurities reduce thermal conductivity by
up to one half by decreasing the phonon mean free path. An even
larger thermal conductivity reduction, with the same physical
origin, occurs in presence of structural defects including
vacancies and edges in narrow graphene nanoribbons. Our
calculations reconcile results of former studies, which differed
by up to an order of magnitude, by identifying limitations of
various computational approaches.
\end{abstract}



\maketitle




\section{Introduction}

With increasing performance of microprocessors, rising heat
evolution poses a serious problem.\cite{IEEE-heat2006} To prevent
damage, excess heat is conducted away to a heat sink using
interconnects with high thermal conductivity. In diamond, which is
used for this purpose and which conducts heat by phonons, isotopic
impurities reduce its excellent thermal conduction by up to one
half.\cite{{Wei1993},{Anthony1990}} The initial prediction that
thermal conductivity of perfect carbon nanotubes and graphene
monolayers (not graphite) should be similar and even surpass the
diamond values\cite{DT130} was subsequently confirmed
experimentally, albeit with a large scatter in the observed
values.\cite{Balandin2011} The added benefit of nanotubes and
graphene is their dual role as thermal conductors and active
elements in electronic circuits.
Unlike in heavier elements, the $^{13}$C/$^{12}$C mass ratio does
modify the phonon spectra of graphitic nanostructures
significantly, causing a large reduction in thermal conductivity
of systems with both
isotopes.\cite{{ruoffthermo2012},{Maruyama-isotope-scattering2011}}

To obtain microscopic understanding of factors limiting thermal
conductivity in graphitic nanostructures, we perform large-scale
non-equilibrium molecular dynamics (MD) simulations of defective
carbon nanotubes, graphene and graphene nanoribbons. We determine
the temperature-dependent thermal conductivity of $^{12}$C-based
systems as a function of $^{13}$C concentration and compare the
effect of isotopic impurities to that of divacancies. We show
that, depending on temperature, the thermal conductivity of
$^{13}$C$_{x}$$^{12}$C$_{1-x}$ nanostructures may be quenched by
up to one half with respect to isotopically pure systems. Even at
low concentrations, atomic vacancies quench thermal conductance
more efficiently than isotopic impurities. Whereas freely
suspended graphene monolayers conduct heat almost as well as
isolated carbon nanotubes, edge scattering reduces significantly
the thermal conductivity of graphene nanoribbons. Our calculations
reconcile results of former studies, which differed by up to an
order of magnitude, by identifying limitations of various
computational approaches.

Since in carbon nanostructures the electronic density of states at
the Fermi level is either zero (diamond, graphene) or very small
(nanotubes, graphene nanoribbons), thermal transport in these
systems is dominated by phonons. According to Fourier law, the
thermal conductivity $\lambda$ is given by the heat current
$dQ/dt$ through area $A$ in response to a temperature gradient
$dT/dz$ as
\begin{equation}
\frac{1}{A}\frac{dQ}{dt}=-\lambda\frac{dT}{dz}\;. %
\label{eq1}
\end{equation}
%
%
The phonon component of the thermal conductivity, which is
dominant, is the product $\lambda = (1/3) c_V v_s
{\langle}l{\rangle}$, where $c_V$ is the specific heat per volume,
$v_s$ the speed of sound, and ${\langle}l{\rangle}$ is the phonon
mean free path. Rigid interatomic bonds in both $sp^2$ and $sp^3$
carbon structures are responsible for a very high speed of sound
$v_s$ and hard phonon modes, which translate into a high Debye
frequency and large value of the specific heat $c_V$. In
isotopically pure monocrystalline diamond and carbon nanotubes,
the phonon mean free path ${\langle}l{\rangle}$ may approach a
large fraction of a micrometer, giving rise to record thermal
conductivity values\cite{{Wei1993},{Balandin2011}} as large as
${\approx}40,000$~Wm$^{-1}$K$^{-1}$ near $T{\approx}100$~K.

Since presence of defects, including isotopic impurities and
atomic vacancies of different types,
can not be avoided in realistic systems, it is imperative to
understand their role in thermal conductivity. We expect defects
to play only a minor role in changing the speed of sound and
specific heat, but to reduce drastically the phonon mean free path
and thus the value of $\lambda$. We believe that the large scatter
in the observed data\cite{Balandin2011} comes not only from the
extreme difficulty to measure this quantity in excellent thermal
conductors, but more importantly due to different types and
concentrations of defects in different samples. Since controlling
defects on the nanometer scale is nearly impossible
experimentally, computer simulations provide a welcome alternative
to understand the effect of particular defects on thermal
conductivity.

\section{Method}

To simulate computationally the conduction of heat, we make use of
large-scale non-equilibrium molecular dynamics (NEMD)
simulations,\cite{{Maeda95},{Evans-NEMD},{Hansen1994},{Gil83}}
which had been used successfully to predict thermal conductivity
of nanotubes and graphene.\cite{DT130} The alternative way to
calculate $\lambda$ using direct MD simulations based on
Eq.~(\ref{eq1}) requires applying a thermostat that maintains a
finite temperature difference ${\Delta}T$ across a finite distance
${\Delta}z$. To prevent artifacts, ${\Delta}z$ must be larger than
the phonon mean free path of up to $1~\mu$m, which is
computationally impracticable. A second alternative, which does
not suffer from this limitation,\cite{SchellingPRB02} is based on
the Green-Kubo formula\cite{McQuarrie-StatMech} that relates
$\lambda$ to the time-averaged autocorrelation function
of the heat flux in the system. As shown earlier,\cite{DT130} this
time average converges very slowly in an equilibrium MD simulation
and depends sensitively on the initial conditions, making
extensive ensemble averaging a necessary requirement that is
computationally extremely demanding for systems of interest
here.\cite{ZhangPRB2011} Due to these
problems,\cite{SchellingPRB02} thermal conductivity calculations
of graphitic nanostructures based on direct
MD,\cite{{Maruyama-theory2006},{JiangJAP10}}
NEMD,\cite{Zhang-nnt2004} the Green-Kubo formalism,\cite{{Grujicic2005},%
{ZhangPRB2011},{ZhangJN2010}} or the Landauer non-equilibrium
Green's function formalism\cite{{SavicPRL08},{WangAPL11}} have
arrived at inconsistent results that differed by up to an order of
magnitude and thus need to be revisited.

Our computational approach\cite{Maeda95} combines the Green-Kubo
formula\cite{McQuarrie-StatMech} with non-equilibrium molecular
dynamics\cite{{Evans-NEMD},{Hansen1994}} in a computationally
efficient manner.\cite{Rapaport-book} The dynamics of the system
is driven by forces
\begin{equation}
\mathbf{F}_i = m_i\frac{d^2\mathbf{r}_i}{dt^2} = %
-\nabla_i U - \zeta m_i \mathbf{v}_i + \Delta\mathbf{F}_i\;,
\label{eq2}
\end{equation}
which act on individual atoms $i$. Here, $\mathbf{r_i}$ is the
position and $\mathbf{v_i}$ the velocity of atom $i$ with mass
$m_i$.

The first term is the gradient of the total potential energy $U$
of the system, representing the force caused by interatomic
interactions. To reduce the unusually high computational
requirements, we represent $U$ by the Tersoff bond-order
potential,\cite{Tersoff88} which reproduces well the optimum
structure as well as vibration spectra of graphene and related
nanostructures.\cite{thermo12-EPAPS} This potential also formally
allows a decomposition of the total potential energy into
potential energies $u_i$ of individual atoms, $U=\sum_i u_i$,
which will be of use in the following.


The second term describes a Nos\'{e}-Hoover
thermostat\cite{Nose,Hoover} that represents coupling of the
system to a heat bath at temperature $T$. The dynamics of the
generalized coordinate $\zeta$ of the heat bath is governed by
\begin{equation}
\frac{d\zeta}{dt} = %
\frac{1}{Q}\left(E_K - \frac{3N-6}{2}k_{B}T\right)\;, %
\label{eq3}
\end{equation}
where $E_K$ is the kinetic energy of the $N$-atom system and $Q$
the thermal inertia of the heat bath.

\begin{figure*}[!tb]
\includegraphics[width=2.0\columnwidth]{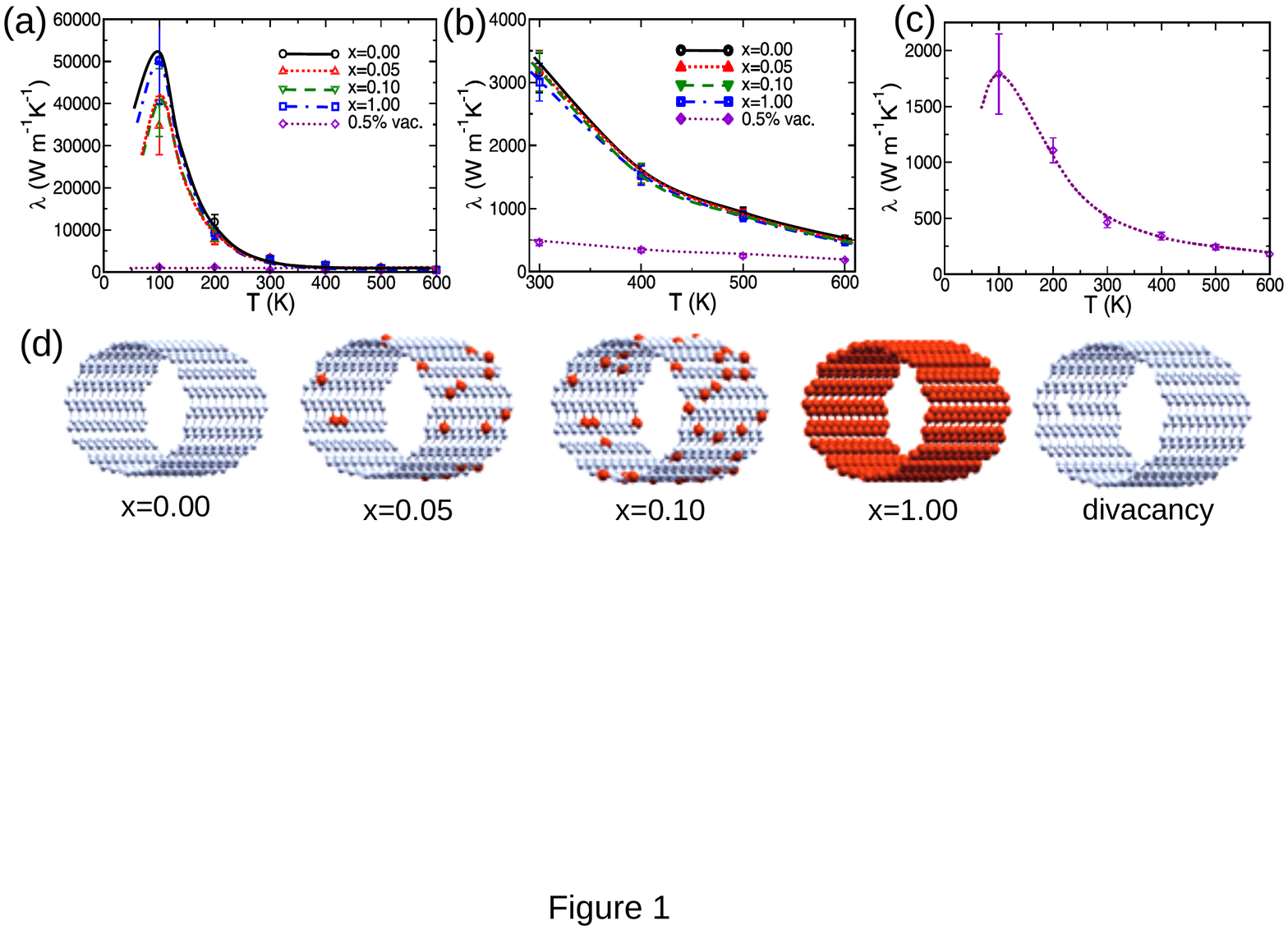}
\caption{(Color online) Thermal conductivity $\lambda$ of perfect
and defective $(10,10)$ carbon nanotubes as a function of
temperature $T$. (a) $\lambda$ in structurally perfect
$^{13}$C$_{x}$$^{12}$C$_{1-x}$ nanotubes with varying isotopic
composition in comparison to pure $^{12}$C nanotubes with 0.5\%
missing atoms forming divacancy defects. (b) Details of (a) on a
reduced temperature scale. (c) $\lambda$ of $^{12}$C nanotubes
with divacancies, presented in (a) and (b), on an expanded
$\lambda$ scale. (d) Depiction of the nanotube unit cell
containing different types of defects. The lines in (a)--(c) are
guides to the eye.
}%
\label{fig1}
\end{figure*}

The third term is a small fictitious force that acts as a
perturbation, driving the system out of equilibrium to generate a
heat flux, and is given by
\begin{equation}
\Delta\mathbf{F}_i={\Delta}e_i \mathbf{F}_e -
                  \sum_{j{\ne}i} \mathbf{f}_{ij}
                  (\mathbf{r}_{ij}{\cdot}\mathbf{F}_e) +
                  \frac{1}{N}\sum_{j}\sum_{k{\ne}j} \mathbf{f}_{jk}
                  (\mathbf{r}_{jk}{\cdot}\mathbf{F}_e)\;.
\label{eq4}
\end{equation}
Here, $\mathbf{F}_e$ is a vector parameter (with the dimension of
inverse length) representing the strength of the perturbation.
$\mathbf{r}_{ij}= \mathbf{r}_{j}-\mathbf{r}_{i}$ and ${\Delta}e_i=
e_i-{\langle}e{\rangle}$ is the excess energy of atom $i$, where
$e_i=m_iv_i^2/2+u_i$ is its instantaneous energy and
${\langle}e{\rangle}=1/N\sum_i e_i$.
$\mathbf{f}_{ij}=-{\nabla_i}u_j$ represents the contribution to
the force on atom $i$ stemming from its interaction with atom $j$,
where $\nabla_i$ is the gradient with respect to the position of
atom $i$.

The heat flux in the system is then given by
\begin{equation}
\mathbf{J}(t)=\frac{d}{dt}\sum_{i=1}^N \mathbf{r_i}{\Delta}e_i =
              \sum_i \mathbf{v}_i{\Delta}e_i -
              \sum_i\sum_{j{\ne}i} \mathbf{r}_{ij}
              (\mathbf{f}_{ij}{\cdot}\mathbf{v}_{i})\;.
\label{eq5}
\end{equation}

Setting $\mathbf{F}_e=F_e\hat\mathbf{z}$, the $z$-component of the
heat flux can be obtained using the simplified
expression\cite{{Evans-NEMD},{Gil83}}
\begin{equation}
J_z=\frac{1}{F_e}\sum_i\mathbf{v}_i\Delta\mathbf{F}_i\;,
\label{eq6}
\end{equation}
which leads to the thermal conductivity $\lambda$ along the
$z$-direction
\begin{equation}
\lambda = \lim_{\mathbf{F}_e{\rightarrow}0}
          \lim_{t\rightarrow\infty}
          \frac{\langle{J_z(\mathbf{F}_e,t)}\rangle_t}{F_eTV}\;.
\label{eq7}
\end{equation}
This approach to determine $\lambda$ has been shown to be
equivalent to that obtained using the Green-Kubo
formula,\cite{Evans-NEMD,Gil83} yet is computationally much less
demanding.\cite{NEMD-efficiency}

To determine $\lambda$ in defect-free and defective nanotubes,
graphene and graphene nanoribbons, we integrated the equations of
motion using ${\Delta}t=0.2$~fs as time step. We used the
fifth-order predictor-corrector
algorithm\cite{Gear-predictor-corrector} for Eq.~(\ref{eq2}) and
the fourth-order algorithm to integrate the coupled
Eq.~(\ref{eq3}). We used $Q=10$~a.m.u.$\cdot${\AA}$^2$ for the
thermal inertial of the thermostat, which allowed for efficient
thermalization while not disturbing significantly the dynamics of
the system. The number of time steps needed for a reliable time
average of the heat flux in Eq.~(\ref{eq7}) varied depending on
the system, the temperature, and the value of $\mathbf{F}_e$. We
found that only 500,000 time steps were sufficient to reach
convergence for $F_e>10^{-3}$~{\AA}$^{-1}$, but for smaller values
$10^{-5}$~{\AA}$^{-1}<F_e<10^{-3}$~{\AA}$^{-1}$ we used up to
$2{\times}10^6$ time steps covering a 0.4~ns time period.
The estimated 10-20\% error in the extrapolation of our results
towards $\mathbf{F}_e{\rightarrow}0$ is shown by the error bars of
$\lambda$ in Figs.~\ref{fig1}-\ref{fig3}.

\section{Results and discussion}

\subsection{Thermal conductivity of carbon nanotubes}

Our results for the thermal conductivity of perfect and defective
$(10,10)$ carbon nanotubes are presented in Fig.~\ref{fig1}. We
used periodic boundary conditions with a large unit cell
containing 400 C atoms, depicted in Fig.~\ref{fig1}(d). Focussing
first on structurally perfect nanotubes with the isotopic
composition $^{13}$C$_{x}$$^{12}$C$_{1-x}$, we present in
Figs.~\ref{fig1}(a) and \ref{fig1}(b) the thermal conductivity of
isotopically pure ($^{12}$C and $^{13}$C) nanotubes and isotope
mixtures with $x=0.05$ and $x=0.10$ in the temperature range up to
$600$~K. In all systems, the low-temperature behavior of $\lambda$
is dominated by that of $c_V$, which, along with $\lambda$,
approaches zero for $T{\rightarrow}0$~K and then gradually
increases with increasing temperature. After reaching its maximum,
which occurs near $T{\approx}100$~K in nanotubes, $\lambda$
decreases again due to the decreasing phonon mean free path,
caused by increasing structural disorder at high temperatures. We
find the thermal conductivity to be highest in isotopically pure
$^{12}$C nanotubes, with isotopically pure $^{13}$C nanotubes
reaching almost the same value. We represented nanotubes with an
isotopic mixture $x=0.05$ and $x=0.10$ by randomly distributing
$^{13}$C atoms across the $^{12}$C lattice. Our results for these
mixtures indicate that for $T<300$~K the thermal conductivity may
decrease by up to 30\% with respect to the value in isotopically
pure lattices due to the strong reduction of the phonon mean free
path.\cite{SavicPRL08} Close inspection of our results reveals
that thermal conductivity of nanotubes remains almost unaffected
by the presence of $^{13}$C isotopic impurities at temperatures
$T{\agt}300$~K, where phonon-phonon scattering seems to dominate
the mean free path reduction.

To find out the relative importance of isotopic impurities and
structural defects, we also studied thermal conductivity in
isotopically pure $^{12}$C $(10,10)$ nanotubes containing a small
fraction of atomic vacancies. We focussed on divacancies, which
are more stable than monatomic
vacancies,\cite{{Banhart-defects2011},{Berber-divacPRB2008}} and
considered one single divacancy per 400-atom unit cell. Our
results for $\lambda$ in this system, shown in
Figs.~\ref{fig1}(a)-\ref{fig1}(c), indicate that even a very low
concentration of structural defects may quench thermal
conductivity by roughly an order of magnitude. These results
support our intuition that vacancies scatter phonons very
efficiently and reduce the phonon mean free path even more than
isotopic impurities. Whereas the reduction of $\lambda$ by 97\% at
$T{\approx}100$~K is extremely large, the relative role of
structural defects decreases at higher temperatures, reaching a
value of 83\% at $300$~K and 67\% at $600$~K. In accord with our
findings for isotope mixtures, we conclude that mean free path
reduction by phonon-phonon scattering starts dominating the
adverse effect of defects at high temperatures.

\subsection{Thermal conductivity of graphene}

\begin{figure*}[!tb]
\includegraphics[width=2.0\columnwidth]{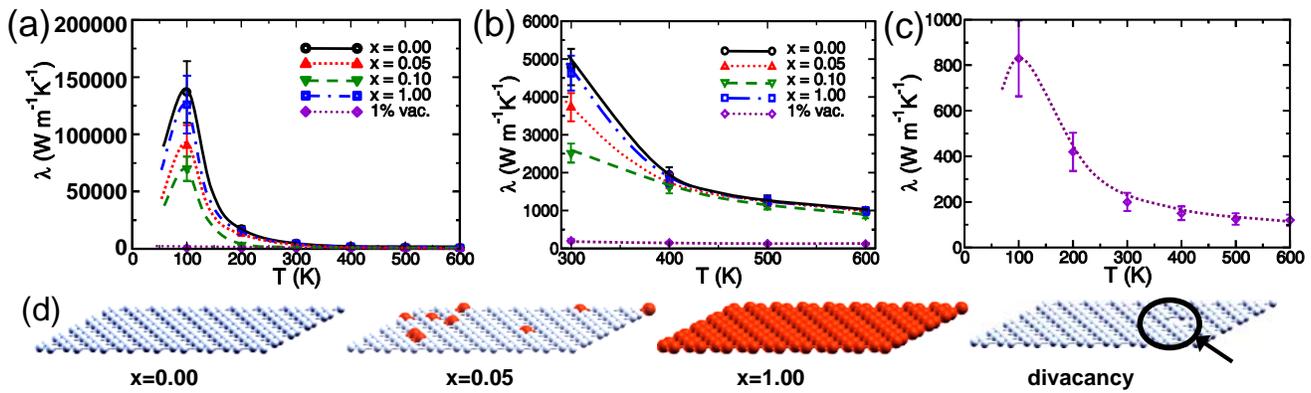}
\caption{(Color online) Thermal conductivity $\lambda$ of perfect
and defective graphene and graphene nanoribbons as a function of
temperature $T$. (a) $\lambda$ in $^{13}$C$_{x}$$^{12}$C$_{1-x}$
graphene with varying isotopic composition in comparison to pure
$^{12}$C graphene with 1\% missing atoms forming divacancy
defects. (b) Details of (a) on a reduced temperature scale. (c)
$\lambda$ of $^{12}$C graphene with divacancies, presented in (a)
and (b), on an expanded $\lambda$ scale. (d) Depiction of the
graphene unit cells containing different types of defects. The
lines in (a)--(c) are guides to the eye.
}%
\label{fig2}
\end{figure*}

Due to the present interest in graphene, we determined the
influence of defects on its thermal conductivity and present our
results in Fig.~\ref{fig2}. Results for $\lambda$ in structurally
perfect, free-standing infinite graphene monolayers with the
isotopic composition $^{13}$C$_{x}$$^{12}$C$_{1-x}$ are shown in
Figs.~\ref{fig2}(a) and (b) for a selected set of compositions in
the temperature range up to $600$~K. Graphene monolayers were
represented by a periodic array of rectangular 180-atom unit
cells. As in the case of nanotubes, isotopically pure graphene has
the highest thermal conductivity.
Similar to nanotubes, the maximum value of $\lambda$ is reached
near $T{\approx}100$~K.

We find the thermal conductivity to be slightly higher in
defect-free graphene consisting of $^{12}$C than of $^{13}$C. In
contrast to nanotubes, the reduction of the thermal conductivity
in isotopic mixtures is much more pronounced in graphene. Whereas
a 10\% content of $^{13}$C isotopic impurities reduced the thermal
conductivity at $100$~K by 30\% in nanotubes, the corresponding
50\% reduction in graphene is much larger. This is consistent with
the fact that in defect-free systems, phonon-phonon scattering
limits the phonon mean free path more in nanotubes with a finite
circumference than in graphene.

Similar to nanotubes, the presence of a single divacancy per unit
cell quenches the thermal conductivity by more than an order of
magnitude, as seen in Figs.~\ref{fig2}(a)--\ref{fig2}(c). Also in
graphene, the reduction of the thermal conductivity due to
isotopic and structural defects is mainly caused by the decrease
in the phonon mean free path.

\subsection{Thermal conductivity of graphene nanorribons}

\begin{figure}[!b]
\includegraphics[width=1.0\columnwidth]{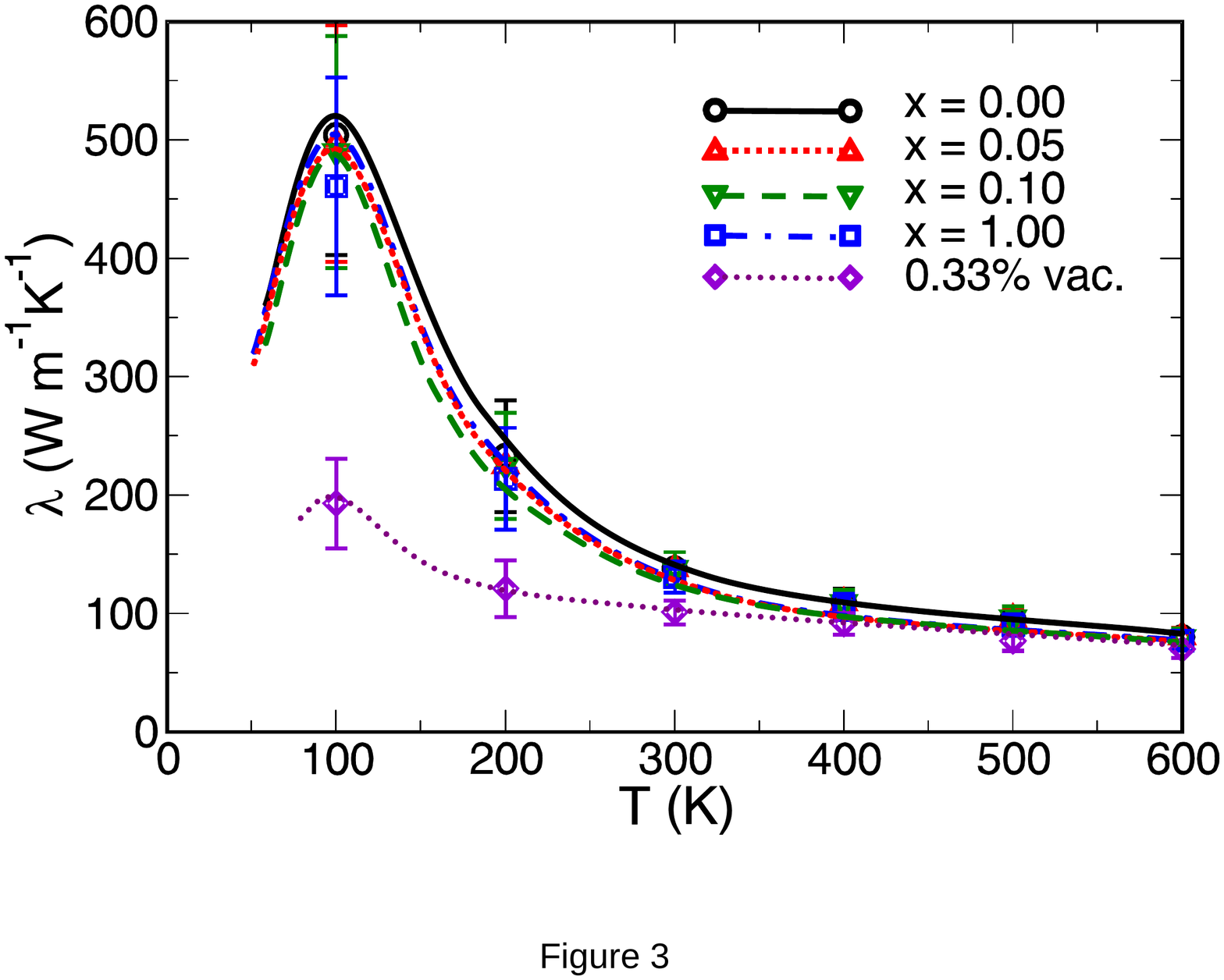}
\caption{(Color online) Temperature dependence of the thermal
conductivity $\lambda$ of $^{13}$C$_{x}$$^{12}$C$_{1-x}$
structurally perfect graphene nanoribbons in comparison to pure
$^{12}$C graphene nanoribbons with 0.33\% missing atoms arranged
as divacancies. The lines are guides to the eye.
}%
\label{fig3}
\end{figure}

Finally, we compared thermal conductivity of graphene to that of
$11.1$~{\AA} wide armchair graphene nanoribbons, using the same
$^{13}$C$_{x}$$^{12}$C$_{1-x}$ isotopic compositions as for
graphene and nanotubes.
The nanoribbons were represented using periodic boundary
conditions using 131.5~{\AA} long unit cells containing 600 C
atoms. Our results for the thermal conductivity of nanoribbons are
presented in Fig.~\ref{fig3}.

Our results for finite-width nanoribbons should be very relevant
also for graphene formed by chemical vapor deposition (CVD). Grain
boundaries in CVD graphene, viewed as lines of incorrectly
coordinated carbon atoms, will scatter phonons and limit thermal
conductivity in a very similar way as nanoribbon edges. Since the
unit cell size representing realistic grain boundaries is
prohibitively large for atomistic simulations, we only point out
the analogy between nanoribbons and polycrystalline graphene.

For isotopically pure $^{12}$C based systems, comparison between
$\lambda$ of graphene in Fig.~\ref{fig2}(a) and graphene
nanoribbons in Fig.~\ref{fig3} reveals that edges in nanoribbons
quench thermal conductivity in a much more drastic way than sparse
diatomic vacancies in an infinite graphene monolayer. We observe
reduction of $\lambda$ by a factor of $700$ at $T=100$~K, a factor
of $30$ at $300$~K, and by an order of magnitude at $600K$.

Possibly unexpected at the first glance is our finding that
placing one divacancy per 600-atom unit cell, corresponding to
0.33\% atomic vacancies, does not cause as drastic a reduction of
the thermal conductivity as in the case of nanotubes and graphene.
Whereas this effect is still significant at low temperatures,
amounting to a 60\% reduction at $T=100$~K, it becomes negligibly
small at temperatures above $400$~K. We conclude that the
underlying reduction of the phonon mean free path in narrow
graphene nanoribbons is dominated by the presence of edges and
that additional structural defects play only a minor role,
especially at higher temperatures.

Due to the dominating role of edges as scattering centers in
nanoribbons, the role of isotopic impurities is much smaller in
these systems than in nanotubes and graphene. Still, in agreement
with our results for nanotubes and graphene, we find that
isotopically pure nanoribbons based on $^{12}$C conduct heat
slightly better than those based on $^{13}$C.

\subsection{High-temperature behavior}

Even though infinitely extended graphene appears to conduct heat
better than nanotubes at low temperatures, the difference between
the two systems becomes smaller at $400$~K and above. As already
mentioned above, the adverse effect of defects on the thermal
conductivity of carbon nanostructures becomes less significant at
higher temperatures, when uncorrelated atomic motion reduces the
phonon mean free path even in defect-free systems. In practice, we
find very similar thermal conductivities in isotopically pure
systems and in isotope mixtures at very high temperatures
$T{\agt}600$~K. At still higher temperatures approaching the
melting point, when vacancy production occurs naturally, presence
of additional structural defects should play a negligible role as
well. At those high temperatures, thermal conductivity of
nanotubes and graphene may drop close to that of nanoribbons.

\section{Summary and Conclusions}

In conclusion, we studied the adverse role of defects on the
thermal conductivity of carbon nanotubes, graphene and graphene
nanoribbons using non-equilibrium molecular dynamics simulations.
We found that all defects, including divacancies, extended edges
and isotopic impurities reduce thermal conductivity significantly
in all systems by introducing phonon scattering centers and thus
decreasing the phonon mean free path. We reconciled results of
former studies, which differed by up to an order of magnitude, by
identifying limitations of various computational approaches. We
found that infinite, defect-free graphene should conduct heat
better than any other carbon nanostructure at low temperatures.
For temperatures $T{\alt}400$~K, isotopic impurities were found to
quench the thermal conductivity of graphene more than that of
carbon nanotubes. We found that even sub-percent concentrations of
divacancies reduced the thermal conductivity of all nanocarbons
more than much higher concentrations of isotopic impurities. For
temperatures $T{\alt}400$~K, the adverse effect of divacancies was
found to be more pronounced in graphene than in carbon nanotubes.
Finite-width graphene nanoribbons can be viewed as graphene with
extended vacancies and thus have not only a much lower thermal
conductivity, but also a lower susceptibility to the presence of
additional defects than graphene. At high temperatures, when
anharmonicities in the force field reduce the phonon mean free
path more than defects, we find that thermal conductivity
decreases significantly and that differences between particular
nanocarbons become washed out to a large degree.

\begin{acknowledgments}
We thank Zhen Zhu for generating vibrational spectra of graphene
using the SIESTA code. This work was funded by the National
Science Foundation Cooperative Agreement \#EEC-0832785, titled
``NSEC: Center for High-rate Nanomanufacturing''. Computational
resources have been provided by the Michigan State University High
Performance Computing Center.
\end{acknowledgments}



%

\end{document}